\documentclass[a4paper]{jpconf}
\usepackage{float}
\usepackage{graphicx}
\usepackage{lineno}
\usepackage{cite}
\begin{document}

% \linenumbers
\title{Forward rapidity $\psi(\mathrm{2S})$ meson production in pp, p-Pb and Pb-Pb collisions with ALICE at the LHC}

\author{Marco Leoncino for the ALICE Collaboration}

\address{INFN and University, Via Pietro Giuria 1, Torino, Italy}

\ead{marco.leoncino@cern.ch}

\begin{abstract}
The ALICE Collaboration has studied the inclusive $\psi(\mathrm{2S})$ meson production in pp, p-Pb and Pb-Pb collisions at the CERN LHC.
The $\psi(\mathrm{2S})$ is detected through its decay to a muon pair, using the forward Muon Spectrometer, which covers the pseudo-rapidity
range $-4<\eta<-2.5$. The $\psi(\mathrm{2S})$ production cross sections in pp collisions are presented as a function of rapidity (y) and transverse 
momentum ($p_{\mathrm{T}}$). In p-Pb collisions, $\psi(\mathrm{2S})$ results are compared to the J/$\psi$ ones by the ratio of their production cross 
sections as a function of rapidity, transverse momentum and event activity. The $\psi(\mathrm{2S})$ nuclear modification factor, $R_{\mathrm{pA}}$, is also 
discussed. The results show a $\psi(\mathrm{2S})$ suppression compared to the one observed for the J/$\psi$ meson and are not described by theoretical models 
including cold nuclear matter effects as nuclear shadowing and energy loss. Finally, the preliminary results of $\psi(\mathrm{2S})$ meson production in Pb-Pb 
collisions are shown in two $p_{\mathrm{T}}$ ranges as a function of the collision centrality. 
\end{abstract}

\section{Introduction}
The study of charmonia (bound states of \textit{c} and \textit{$\bar{c}$} quarks), in different collision systems, is the object of intense theoretical
and experimental investigations \cite{qq_studies}. Proton-proton (pp) collisions are fundamental to evaluate the production cross section and to test 
production models. In proton-nucleus (p-A) collisions, several initial and final state effects, related to the presence of cold nuclear matter (shadowing, 
energy loss and nuclear absorption) can influence the observed charmonium yields \cite{Shadowing1,Eloss1}. Finally, in nucleus-nucleus (A-A) collisions,
a deconfined phase of quarks and gluons (QGP) is expected to play an important role on the charmonium production \cite{Matsui_Satz}. Among the charmonium 
the $\psi(\mathrm{2S})$ meson is receiving a lot of attention since it is more weakly-bound than the J/$\psi$ and intriguing results have been already obtained 
at lower collision energies \cite{Psi2SE866}. ALICE data can improve the understanding of $\psi(\mathrm{2S})$ production in hadronic collisions.

\section{ALICE detector and data samples}
The ALICE detector consists of a central barrel dedicated to particle tracking and identification (in the pseudo-rapidity range of $\left|\eta\right|< 0.9$) 
and a forward spectrometer for the detection of muons (in the interval of $-4<\eta<-2.5$). More details about the experimental setup can be found in 
\cite{ALICE_detector}. Charmonium states are detected in the dimuon decay channel using the Muon Spectrometer. The pp analysis is performed in the 
rapidity interval of $-4<y_{\mathrm{lab}}<-2.5$ using a data sample obtained at the center of mass energy of $\sqrt{s}$=7 TeV and corresponding to an integrated luminosity of $L_\mathrm{{int}}^{pp}=1.35\pm 0.07\>\mathrm{pb}^{-1}$ .
In p-Pb collisions data have been collected at $\sqrt{s_{NN}}$=5.02 TeV in two configurations with inverted beam directions, with the following rapidity coverages: 
$-4.46<y_{\mathrm{cms}}<-2.96$ ($L_\mathrm{{int}}^{\mathrm{Pbp}}=5.81\pm 0.18\>\mathrm{nb}^{-1}$, Pb-going direction) at backward rapidity and $2.03<y_{\mathrm{cms}}<3.53$ ($L_\mathrm{{int}}^{\mathrm{pPb}}=5.01\pm 0.19\>\mathrm{nb}^{-1}$, p-going direction) 
at forward rapidity. Finally, the $\psi(\mathrm{2S})$ production in Pb-Pb collisions is studied at $\sqrt{s_{NN}}$=2.76 TeV ($L_\mathrm{{int}}^{\mathrm{PbPb}}=68.8\pm 0.9\>\mathrm{\mu b}^{-1}$) in the rapidity region
of $-4<y_{\mathrm{lab}}<-2.5$.

\section{Results}
The $\psi(\mathrm{2S})$ cross section is obtained as: $\sigma^{\mathrm{\psi(2S)}} = N^{\psi(\mathrm{2S})}/(L_{\mathrm{int}}\cdot \mathrm{BR}_{\mu^{+}\mu^{-}}\cdot \left \langle A\epsilon \right \rangle)\>$, 
where $N^{\psi(\mathrm{2S})}$, the number of reconstructed $\psi(\mathrm{2S})$, is divided by the branching ratio ${BR}_{\mu^{+}\mu^{-}}$, the detector mean 
acceptance times efficiency $\left \langle A\epsilon \right \rangle$ and finally normalized to the integrated luminosity $L_{\mathrm{int}}$.

\vspace{-8pt}

\subsection{pp collisions}
The results in pp collisions \cite{ALICEpp} are shown in Fig.1: the $p_{\mathrm{T}}$-differential cross section is compared to LHCb results \cite{LHCbRes}. A good agreement is observed between the two 
experiments (small differences are visible at low $p_{\mathrm{T}}$, but the comparison is not trivial because of the different rapidity coverage of the
two detectors).

\vspace{-8pt}

\begin{figure}[H]
\begin{center}
\includegraphics[width=0.42\textwidth]{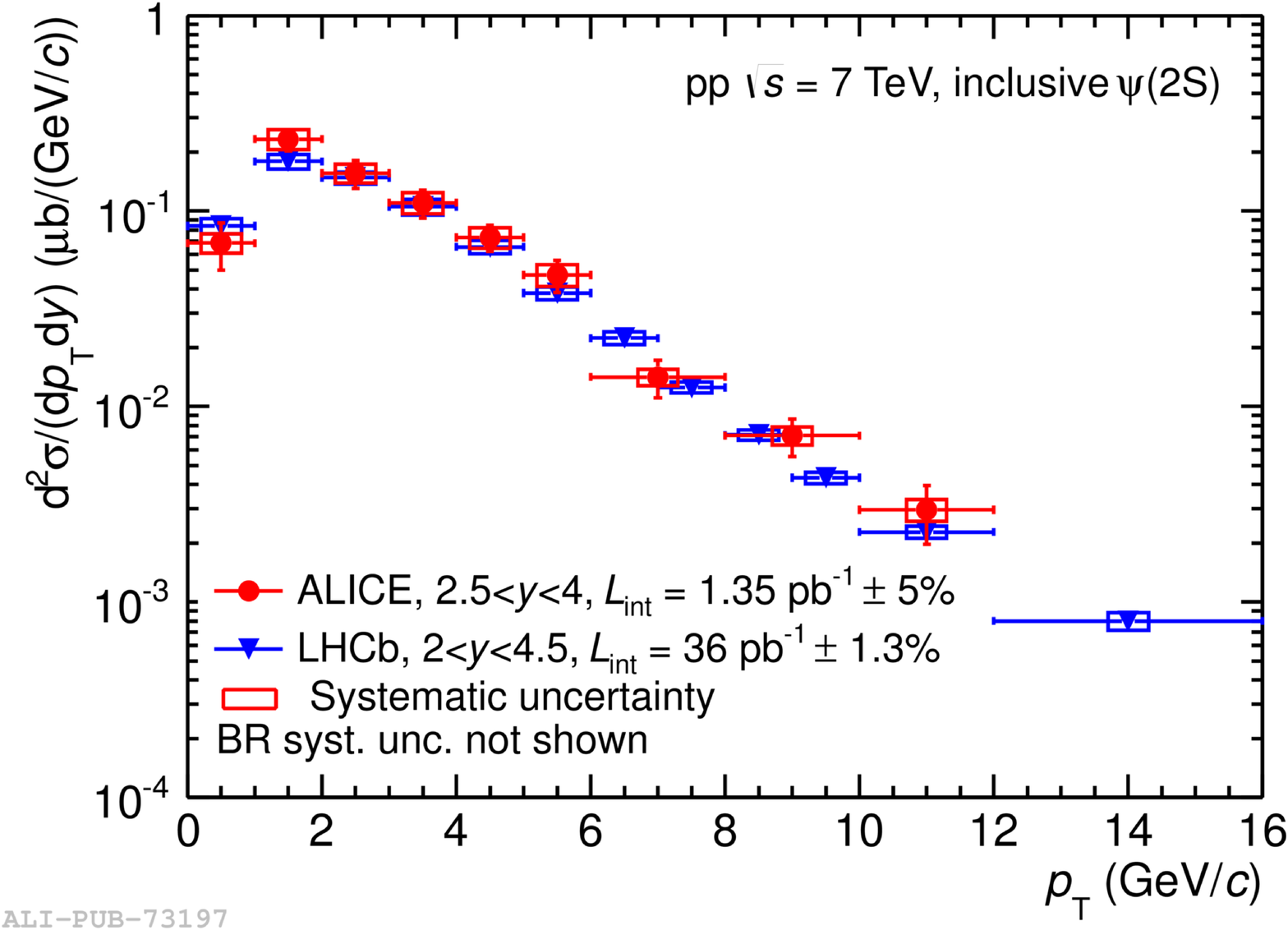}
\qquad
\includegraphics[width=0.42\textwidth]{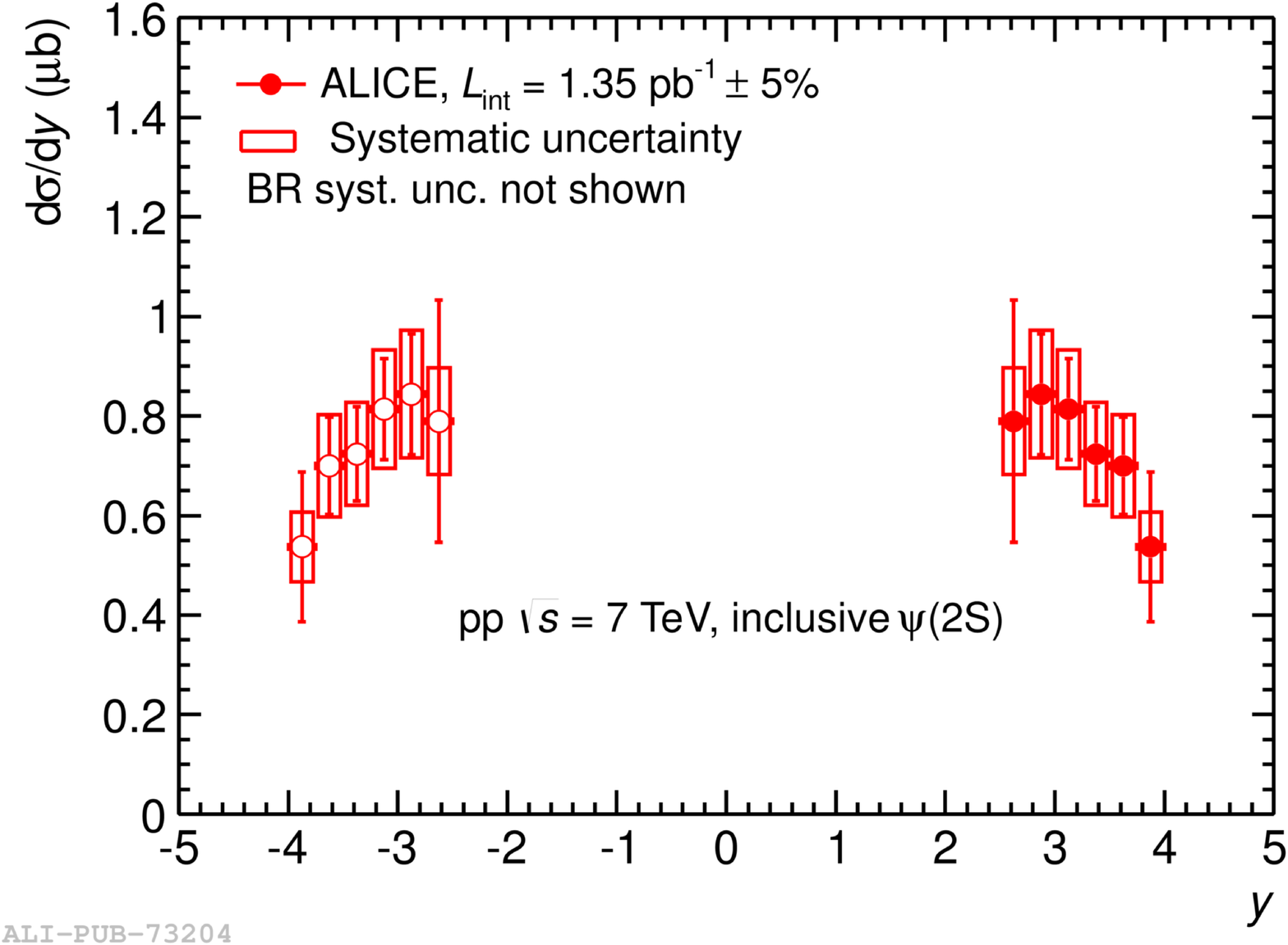}
\caption{Differential $\psi(\mathrm{2S})$ production cross section as a function of $p_{\mathrm{T}}$ (left) and $y$ (right). The $p_{\mathrm{T}}$ differential
results are compared to LHCb measurements \cite{LHCbRes}.}
% The open symbols in the right plot are the reflection of the positive-y measurements with to respect $y=0$.
\end{center}
\end{figure}

\vspace{-30pt}

\subsection{p-Pb collisions}
The cross section ratio between the tightly bound J/$\psi$ and the loosely bound $\psi(\mathrm{2S})$ charmonium states,
${\rm B.R.}_{\psi(\rm 2S)\rightarrow\mu^+\mu^-}\sigma_{\psi(\rm 2S)}/{\rm B.R.}_{{\rm J}/\psi\rightarrow\mu^+\mu^-}\sigma_{{\rm J}/\psi}$
is shown in the left panel of Fig.2. These ratios are significantly lower than the ones in pp, both at forward and backward rapidity, 
pointing to a bigger $\psi(\mathrm{2S})$ suppression (compared to the J/$\psi$) in p-Pb collisions than in pp. 

The double ratios together with that of PHENIX, $[\sigma_{\psi(\rm 2S)}/\sigma_{\rm J/\psi}]_{\rm pPb}/[\sigma_{\psi(\rm 2S)}/\sigma_{\rm J/\psi}]_{\rm pp}$
is shown in the right panel of Fig.2. These results indicate thate the $\psi(\mathrm{2S})$ suppression is more than the J/$\psi$ to a level
of 2.1$\sigma$ at forward-rapidity and 3.5$\sigma$ at backward-rapidity. At midrapidity, PHENIX results \cite{PhenixdAu}, from $\sqrt{s_{\rm NN}}$~=~200 GeV 
d-Au collisions, are in qualitative agreement with ALICE data \cite{ALICE_pPb}.

\begin{figure}[h]
\begin{center}
\includegraphics[width=0.42\textwidth]{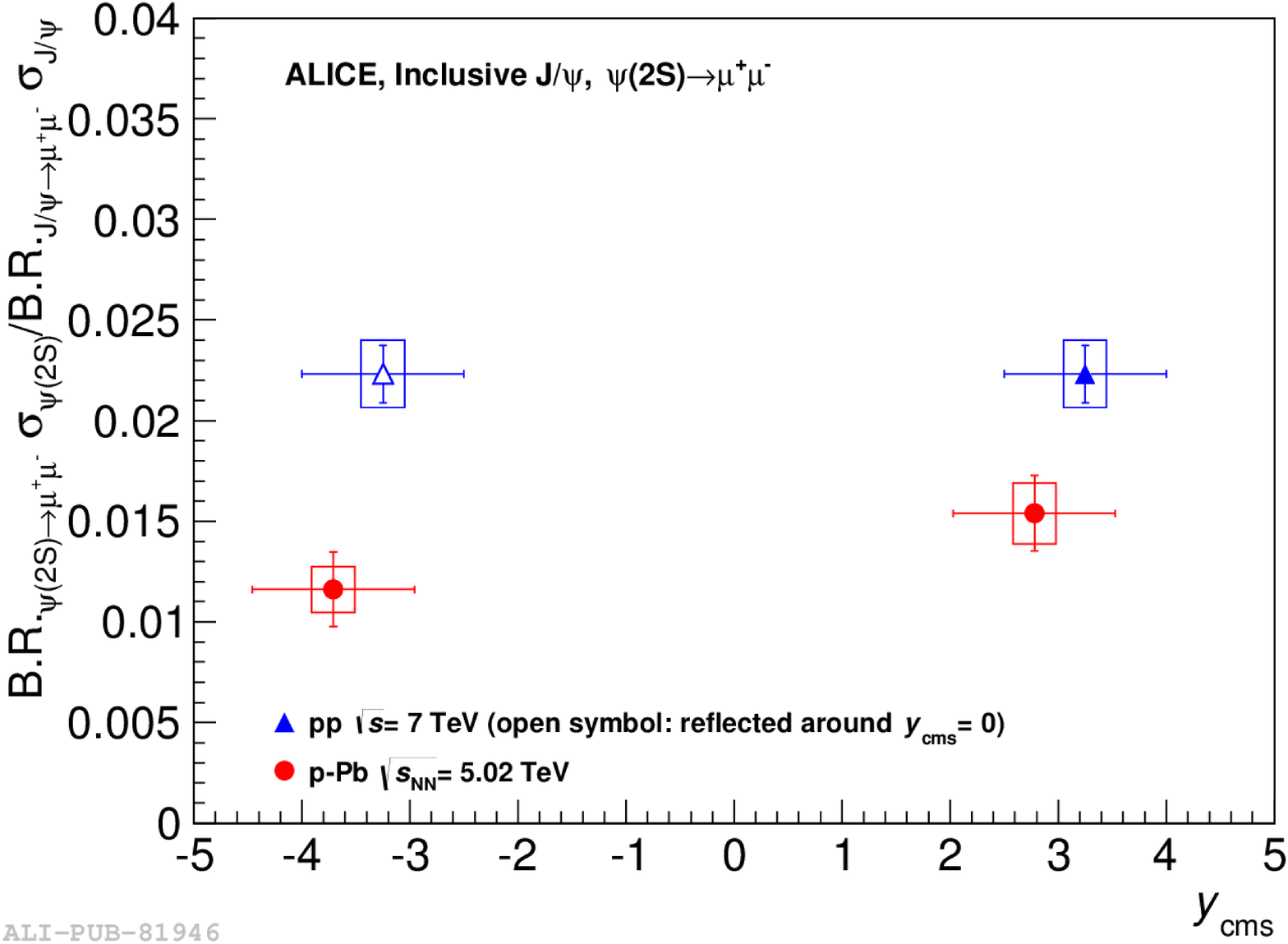}
\qquad
\includegraphics[width=0.42\textwidth]{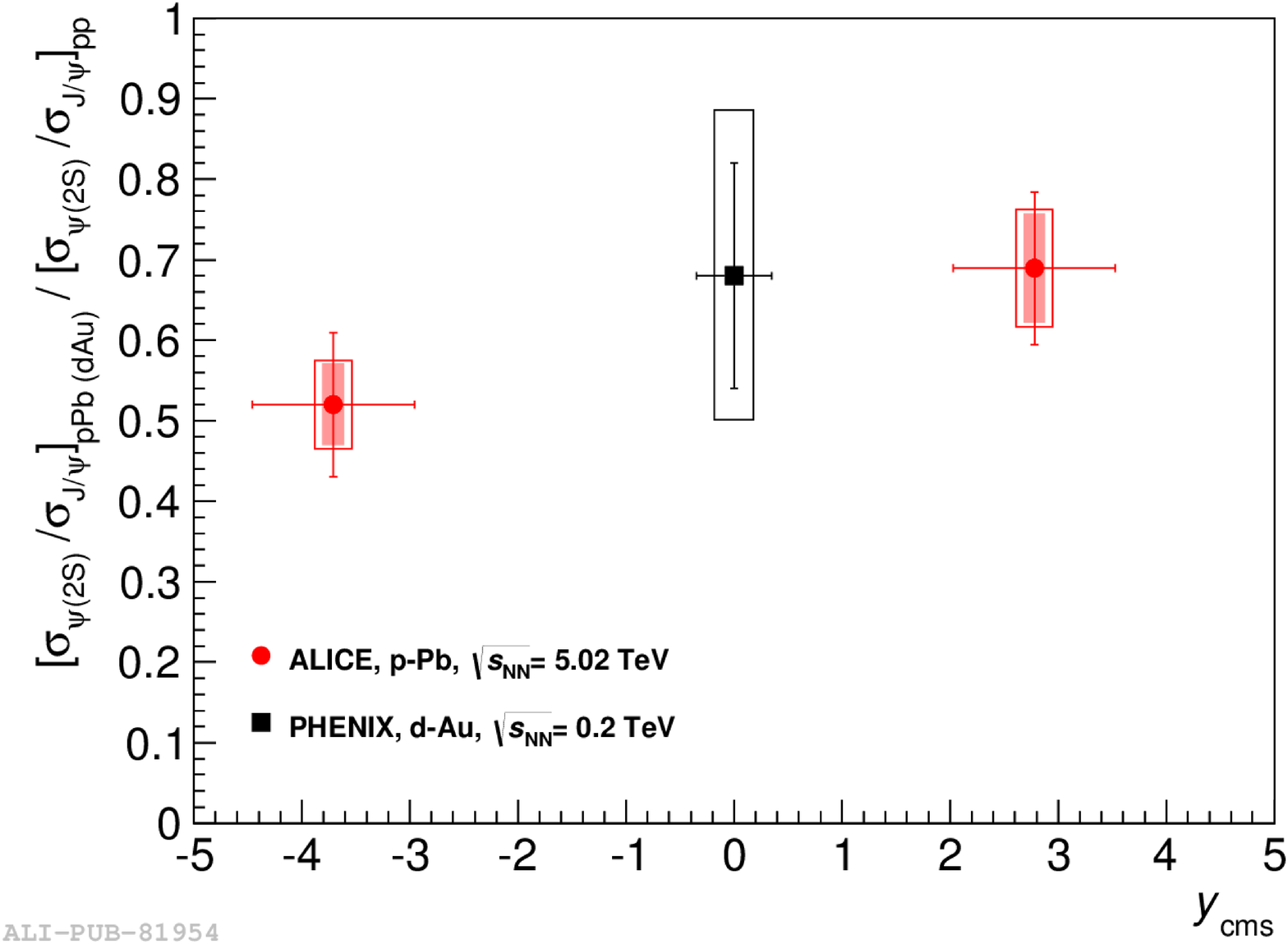}
\caption{Left: the cross section ratios compared with the corresponding pp results at $\sqrt{s}=7$ $\mathrm{TeV}$. Right: the double ratios compared to the corresponding 
PHENIX result \cite{PhenixdAu}.}
\end{center}
\end{figure}

The nuclear modification factor $R_{\mathrm{pA}}$, i.e. the ratio of the $\psi(\mathrm{2S})$ production yield in p-A to the one in pp scaled by the number of binary collisions, 
is another useful quantity to study the effects of nuclear matter on the $\psi(\mathrm{2S})$ production. The $R_{\mathrm{pA}}$ of $\psi(\mathrm{2S})$ and J/$\psi$, are shown 
in Fig.3, left, in the two rapidity intervals, indicating a stronger $\psi(\mathrm{2S})$ suppression than that of the J/$\psi$, both at backward and forward rapidity.

\begin{figure}[H]
\begin{center}
\includegraphics[width=0.42\textwidth]{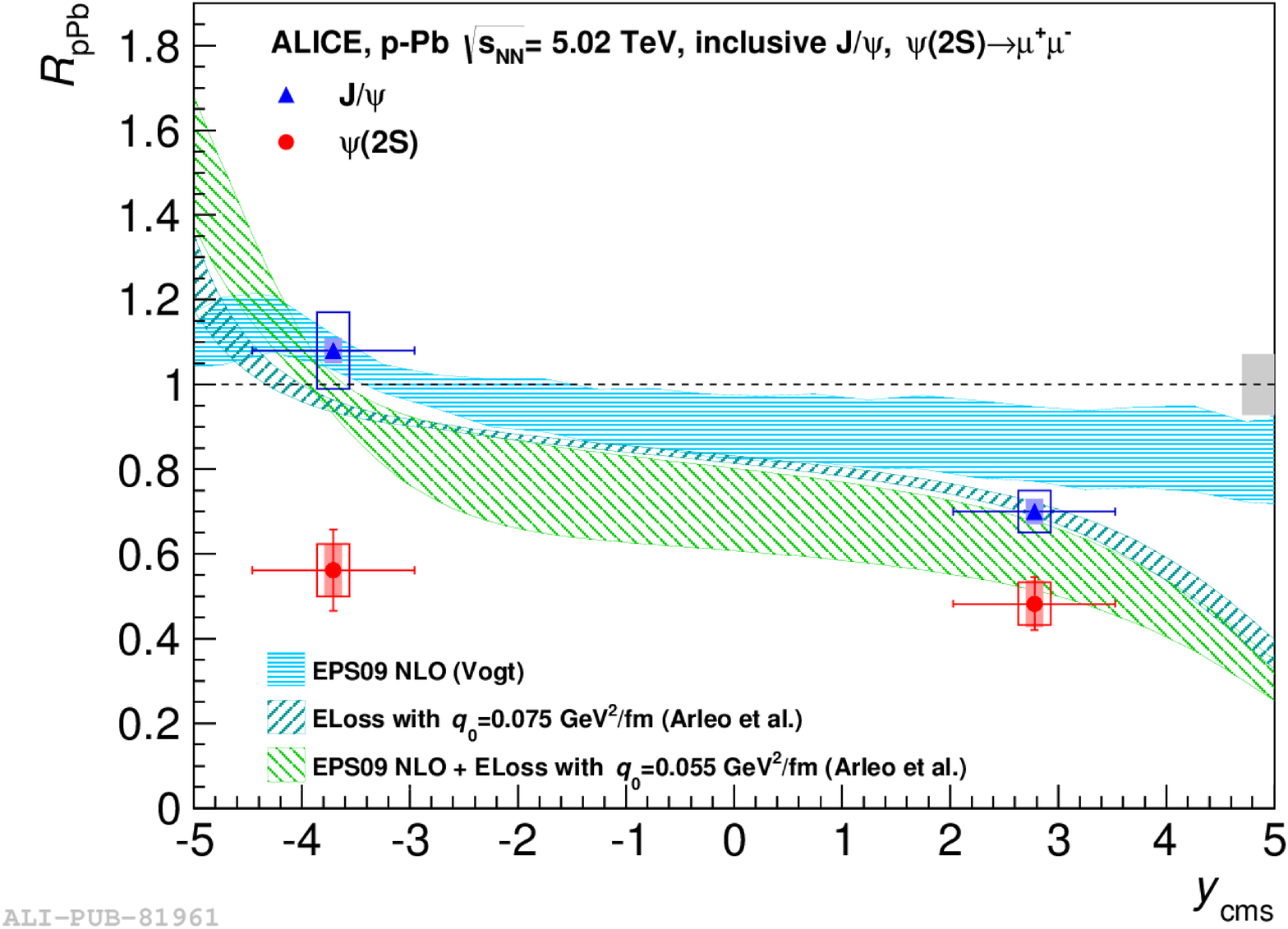}
\qquad
\includegraphics[width=0.42\textwidth]{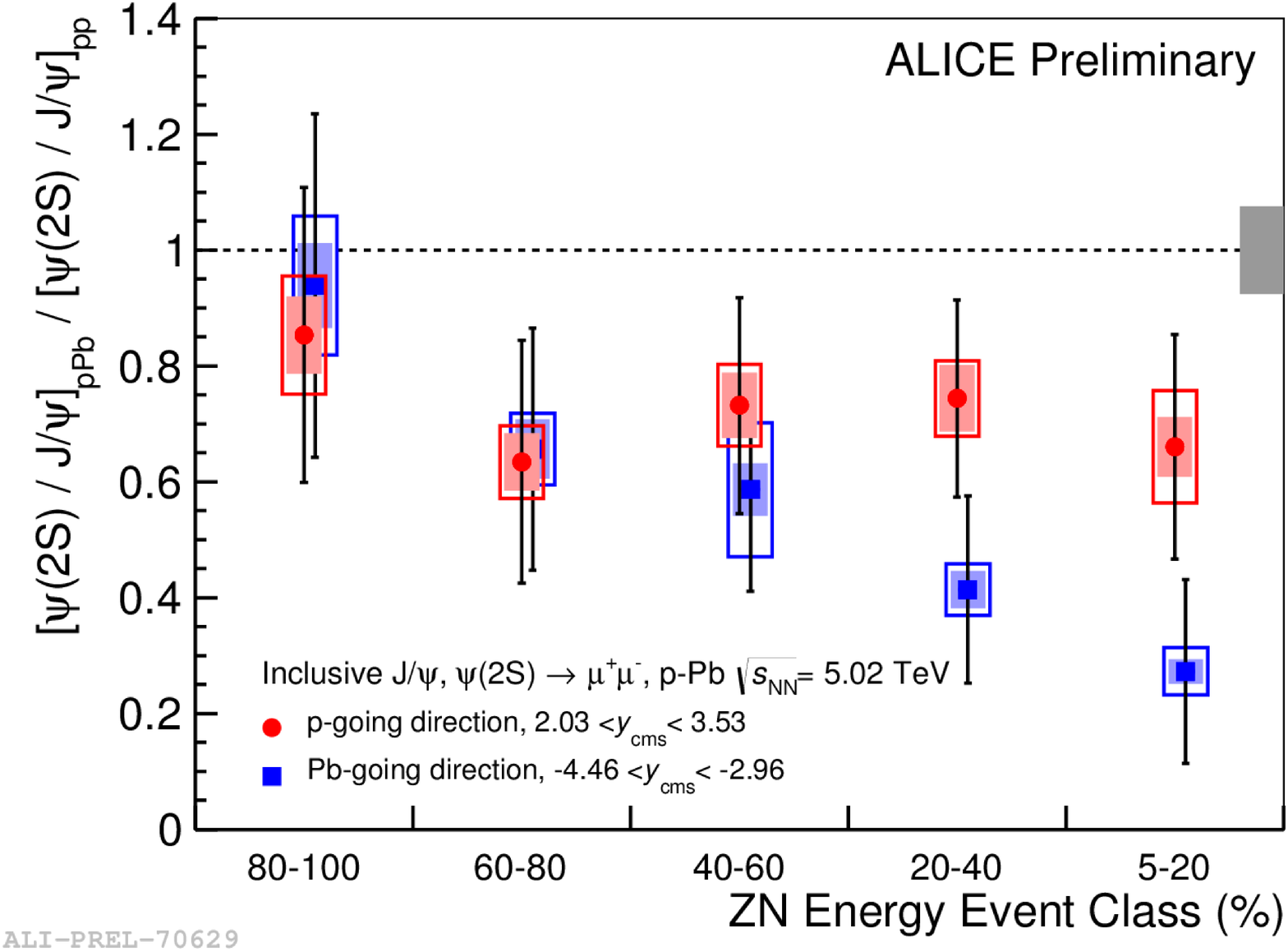}
\caption{Left: the nuclear modification factor for $\psi(\rm 2S)$ compared to the corresponding J/$\psi$ one. Model calculations tuned on J/$\psi$ and including nuclear shadowing and coherent
energy loss are also shown. Right: double ratios as a function of the event activity in p-Pb and Pb-p collisions.}
\end{center}
\end{figure}

\vspace{-20pt}

ALICE results are compared with theoretical predictions including shadowing only \cite{Ramona} or coherent energy loss, with or without a shadowing
contribution \cite{Arleo}. These calculations correspond to the ones performed for the J/$\psi$: shadowing effects are expected to be similar (within 2-3$\%$), 
because of the similar gluon distributions that produce the \textit{c}\textit{$\bar{c}$} state, while no dependence on the final state is expected for coherent energy
loss. The predictions are in disagreement with the $\psi(\rm 2S)$ data and indicate that other final state effects should be considered to explain the observed $\psi(\mathrm{2S})$ 
suppression. The break-up of the resonance in the nuclear medium depends on the binding energy of the charmonium states and could be considered a cause of the larger
$\psi(\mathrm{2S})$ suppression. However, the break-up is relevant only if the charmonium formation time $\tau_{f}$ is smaller than the time $\tau_{c}$ spent by the 
\textit{c}\textit{$\bar{c}$} pair in the nucleus. Estimates for $\tau_{f}$ \cite{tauf1} are in the range 0.05-0.15 $\mathrm{fm/c}$, while $\tau_{c}=\left \langle L \right \rangle/(\beta_{z}\gamma)$ \cite{tauf2} 
(where $\left \langle L \right \rangle$ is the average length of nuclear matter crossed by the pair, $\beta_{z}=\mathrm{tanh}y_{c\bar{c}}^{rest}$ and $\gamma=E_{c\bar{c}}/m_{c\bar{c}}$) 
is about $10^{-4} \mathrm{fm/c}$ at forward rapidity and about $7\cdot10^{-2} \mathrm{fm/c}$ at backward rapidity. In this situation, the strong $\psi(\mathrm{2S})$ suppression 
cannot be explained in terms of the \textit{c}\textit{$\bar{c}$} pair break-up (expecially at backward rapidity where the difference between the J/$\psi$ and
$\psi(\mathrm{2S})$ $R_{\mathrm{pA}}$ is bigger). Finally, the double ratio $[\sigma_{\psi(\rm 2S)}/\sigma_{\rm J/\psi}]_{\rm pPb}/[\sigma_{\psi(\rm 2S)}/\sigma_{\rm J/\psi}]_{\rm pp}$ 
is presented as a function of the event activity (i.e. the event multiplicity based on a measurement from the Zero Degree Calorimeters) in the two rapidity intervals (see Fig.3, right panel). 
When compared to the J/$\psi$, the $\psi(\rm 2S)$ is more suppressed with increasing event activity, in particular at backward rapidity. This could be another hint of final state effects 
that can affect the $\psi(\mathrm{2S})$ production, in particular at backward rapidity.

\subsection{Pb-Pb collisions}
The double ratio $[\sigma_{\psi(\rm 2S)}/\sigma_{\rm J/\psi}]_{\rm PbPb}/[\sigma_{\psi(\rm 2S)}/\sigma_{\rm J/\psi}]_{\rm pp}$ has been studied by ALICE 
as a function of the collision centrality in two $p_{\mathrm{T}}$ intervals (see Fig.4). In the interval $0<p_{\mathrm{T}}<3$ GeV/c, the $\psi(\rm 2S)$ signal 
can be extracted in three centrality classes, while, in the interval $3<p_{\mathrm{T}}<8$ GeV/c the upper limit at 95$\%$ confidence level is shown for the 
most central collisions. ALICE results are compared with the CMS double ratios presented in two $p_{\mathrm{T}}$ intervals corresponding to two different rapidity 
ranges. However, the large statistical and systematic uncertainties of the ALICE results prevent a firm conclusion on the $\psi(\mathrm{2S})$ behaviour in Pb-Pb 
and the comparison with the CMS values \cite{CMSPbPb} is not straightforward, given also the different kinematic coverage.

\begin{figure}[H]
\begin{center}
\includegraphics[width=0.43\textwidth]{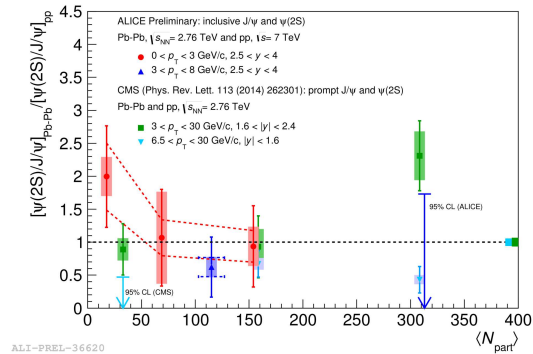}
\caption{Double ratios $[\sigma_{\psi(\rm 2S)}/\sigma_{\rm J/\psi}]_{\rm PbPb}/[\sigma_{\psi(\rm 2S)}/\sigma_{\rm J/\psi}]_{\rm pp}$ as a function of the event 
centrality, in two $p_{\mathrm{T}}$ intervals. CMS measurements \cite{CMSPbPb}, in two $p_{\mathrm{T}}$ intervals corresponding to two different rapidity coverages,
are also shown.} 
\end{center}
\end{figure}

\vspace{-33pt}

\section{Conclusions}
In summary, ALICE collaboration has studied the $\psi(\rm 2S)$ production in pp, p-Pb and Pb-Pb collisions. In pp collisions the $\psi(\rm 2S)$ production
cross sections have been obtained as a function of $p_{\mathrm{T}}$ and $y$, and are in good agreement with the LHCb measurements. In p-Pb collisions the $\psi(\rm 2S)$
is more suppressed than the J/$\psi$ at both forward and backward rapidity. Theoretical models based on shadowing and/or energy loss are in disagreement with data
and the break-up of the \textit{c}\textit{$\bar{c}$} pair can hardly explain the strong $\psi(\rm 2S)$ suppression, indicating that other final state effects
are required. Finally, preliminary results in Pb-Pb collisions have been shown: large uncertainities prevent to make definitive conclusions.

% \item{qq_studies} Strite S and Morkoc H 1992 {\it J. Vac. Sci. Technol.} B {\bf 10} 1237 EXAMPLE

\end{document}